\documentclass[12pt]{iopart}

\usepackage{latexsym}
\usepackage{graphics,epsfig} 

\usepackage{citesort}
\usepackage[square,sort&compress]{natbib}
\usepackage{epstopdf}

\usepackage{iopams}

\begin{document}
\title[Radiative lifetime measurements of rubidium 
Rydberg states]{Radiative lifetime measurements of rubidium 
Rydberg states}

\author{D B Branden$^1$\footnote{Present address: Oliver Wyman LLC, Boston, MA 02116, USA} T Juhasz$^1$\footnote{Present address: Peabody Veterans Memorial High School, Peabody, MA 01960, USA}, 
T Mahlokozera$^1$ \footnote{Present address:
Children's Hospital Boston, Boston MA 02115, USA} C Vesa$^1$\footnote{Present address:
Department of Physics, University of Alberta, Edmonton, T6G 2G7, Canada} R O Wilson$^1$\footnote{Present address:
Blair Academy, Blairstown, NJ 07825, USA} M Zheng$^1$\footnote{Present address:
Department of Physics, University of Illinois, Urbana, IL 61801, USA} A Kortyna$^2$ and D A Tate$^1$}

\address{$^1$ Department of Physics and Astronomy, Colby College, Waterville, ME 04901, USA}
\address{$^2$ Department of Physics, Lafayette College, Easton, PA 18042, USA}

\ead{datate@colby.edu}
\begin{abstract}
We have measured the radiative lifetimes of $ns$, $np$ and $nd$ Rydberg states of rubidium in the range $28 \le n \le 45$. To enable long-lived states to be measured, our experiment uses slow-moving ($\sim$100 $\mu$K) $^{85}$Rb atoms in a magneto-optical trap (MOT). Two experimental techniques have been adopted to reduce random and systematic errors. First, a narrow-bandwidth pulsed laser is used to excite the target $n\ell$ Rydberg state, resulting in minimal shot-to-shot variation in the initial state population. Second, we monitor the target state population as a function of time delay from the laser pulse using a short-duration, millimetre-wave pulse that is resonant with a one- or two-photon transition to a higher energy ``monitor state'', $n^\prime\ell^\prime$. We then selectively field ionize the monitor state, and detect the resulting electrons with a micro-channel plate. This signal is an accurate mirror of the $n\ell$ target state population, and is uncontaminated by contributions from other states which are populated by black body radiation. Our results are generally consistent with other recent experimental results obtained using a method which is more prone to systematic error, and are also in excellent agreement with theory. 

\end{abstract}

\submitto{\JPB}

\maketitle

\section{Introduction}\label{intro}

In the 1970's and early 1980's, there was a significant amount of interest in the radiative lifetimes of Rydberg states of alkali metal atoms. From the experimental perspective, alkali atoms are relatively easy to work with. High atomic densities can be achieved easily in a cell or atomic beam, and the wavelengths needed to photo-excite specific Rydberg states coincide well with those available from the (then) newly-invented tunable, pulsed dye lasers \cite{hans72,litt78}. The state of interest (termed the ``target state'' in this paper) could thus be excited with a temporal precision of 10 ns or less, and the radiative decay could be monitored by observing the emitted fluorescence using a photomultiplier. There is theoretical interest in radiative lifetime measurements of Rydberg states of alkalis since their wave functions are very similar to those of hydrogen, and calculation of atomic properties of alkalis is much less computationally intensive than for other groups of elements. Experimental values for radiative lifetimes provide useful data for testing the accuracy of the wave functions used in the calculations \cite{goun79,theo84,he90,bet09}.

A comprehensive list of Rydberg state lifetime measurements in Li, Na, K, Rb, and Cs as it existed in 1984 can be found in Theodosiou's theoretical paper \cite{theo84}. A very significant issue that became apparent as higher $n$-states (where $n$ is the principal quantum number) were measured was the effect of black body radiation (BBR) on Rydberg state lifetimes. BBR has a negligible effect on low-lying energy states, whose radiative lifetimes are dominated by spontaneous electric dipole transitions to the lowest atomic states allowed by the selection rules. However, dipole matrix elements for $n,\ell$ $\rightarrow$ $n \pm 1,\ell \pm 1$ transitions increase approximately as $n^2$. In addition, the energy spacing of states decreases as $n^{-3}$, and thus the photon energy needed to drive transitions between Rydberg states tends towards the peak of the BBR spectrum as $n$ increases. Above $n \approx 10$, therefore, BBR stimulated depopulation becomes non-negligible compared to the spontaneous decay of a state. Experimentally, the effect of BBR became apparent when the measured Rydberg state lifetimes were found to be significantly lower than those calculated assuming purely spontaneous decay, i.e., ignoring stimulated depopulation by BBR. The ability of BBR to significantly affect Rydberg state lifetimes of alkalis was investigated in a number of experimental studies in the late 1970's \cite{gall79,duca79,beit79}. In particular, Spencer \etal made a thorough study of the effect of changing the BBR spectral temperature on the state lifetime \cite{spen82a}. Calculations of BBR redistribution rates, and the effect of BBR on Rydberg state lifetimes, were made by Farley and Wing \cite{farl81} and Theodosiou \cite{theo84}. Additionally, Galvez \etal \cite{galv99} compared the experimentally observed population redistribution due to BBR stimulated transitions with calculated distributions based on theoretical rate values obtained by similar methods to those reported in \cite{theo84} and \cite{farl81}.

Measurements of Rydberg state lifetimes that use thermal samples are limited to states below $n \approx 20$. This is because the radiative lifetimes increase with $n$, and at $n = 20$, they are of order $10 \ \mu$s. During this time, thermal atoms travel distances on the order of 1 cm. Accurate measurements require observation for several lifetimes, which necessitates a significant detection volume. Additionally, collisional depopulation effects and superradiance become significant at long radiative lifetimes. 

In the past 10-15 years, there has been great interest in cold Rydberg atoms. When made in sufficiently dense samples, they exhibit many-body collision behavior \cite{and98,mou98}, as well as suppressed laser excitation probability due to an interaction-induced blockade (for instance the van der Waals blockade, or the dipole blockade, or some other density-dependent excitation suppression mechanism) \cite{tong04,sing04,lieb05,vogt06}. In addition, novel long-range molecules can be created \cite{stan06,bend09}, and dense, cold Rydberg samples have also been found to evolve spontaneously into ultra-cold, strongly coupled plasma \cite{li04,li05,li06}. All of these studies start with cold atoms in a magneto-optical trap (MOT), and such a source is ideal for measurements of Rydberg atom properties such as quantum defects \cite{li03,han06} and radiative lifetimes \cite{maga00,deoliv02,nasc06,marc09,gabb06,tret09a,feng09}. In the ultra-cold MOT environment, the atomic velocities are less than 1 m/s and hence the atoms remain in the detection volume long enough for an accurate determination of the properties of Rydberg states to be made.

While various atoms have been trapped in a MOT successfully, Rydberg lifetime measurements using cold atoms have so far been reported only for Rb \cite{maga00,deoliv02,nasc06,marc09,gabb06,tret09a} and Cs \cite{feng09}.  In all of these experiments except \cite{gabb06}, the technique that was used to monitor the decay of the Rydberg state population was selective field ionization (SFI) with a $\sim 1 \  \mu$s rise-time high-voltage pulse. In \cite{gabb06}, the technique for monitoring the Rydberg population was to detect photo-electrons or photo-ions created when a mid-infrared ($\approx10 \ \mu$m wavelength) laser pulse illuminated the sample at a well-determined time after the target state was excited. (These techniques are discussed more fully below in section \ref{general}.) These recent experiments have stimulated renewed interest in extending theoretical calculations to the previously unstudied region $n \ge 30$ \cite{bet09}. Additionally, lifetime measurements have been used to distinguish exotic Rydberg molecular states from their parent atomic states \cite{bend09}.

The SFI detection technique that was used to monitor the Rydberg state population in \cite{maga00,deoliv02,nasc06,marc09,tret09a,feng09} must be applied carefully to avoid significant systematic errors in the lifetime measurement. (Photoionization by an infrared laser pulse, the technique used in \cite{gabb06}, may also lead to systematic errors.) Specifically, the arrival times of electrons (or ions) created by SFI at the detector must be carefully gated so that only those from field-ionization of the target state are observed. Due to BBR, states with similar energies to the target are populated by stimulated transitions during the decay time of the target, and if the SFI signal from these nearby states cannot be rejected, the method yields an inaccurate lifetime measurement. One of us has commented on this issue \cite{tate07} with regard to the results of a group at Universidade de S\~{a}o Paulo, which are presented in \cite{maga00,deoliv02,nasc06}, and the comment describes fully the problem and its implications. In their response \cite{cali07}, the S\~{a}o Paulo group clarifies their experimental procedure, but as they observe in \cite{marc09} and \cite{cali07}, systematic uncertainty when the SFI detection technique is used is still possible. 

The present paper presents Rydberg state lifetime measurements of Rb for most of the same states previously reported in \cite{maga00,deoliv02,nasc06,marc09}. However, our measurements were made with a modification of the SFI technique that allowed the decay of the target state to be unambiguously distinguished from signals arising from the effects of BBR stimulated transfer of population to other Rydberg states which also contribute to the SFI signal. This was achieved by using frequency-tuned pulses of millimetre-waves (mm-waves) to sample the population of the target state as a function of delay after the exciting laser. Previous work by a group at University of Virginia \cite{li03, han06} has determined the frequencies of $n\ell$ $\rightarrow$ $n^\prime \ell^\prime$ mm-wave transitions in Rb with great precision, and we use this knowledge to accurately identify which part of the SFI signal is characteristic of the target state population.

Below, we describe our experiment and our results. First, however, we discuss the present state of lifetime calculations. Theoretical considerations have an important impact on this experiment, since our lifetime measurements are made in a BBR field with a spectral temperature of nominally 300 K. While our experiment (as with the previous cold-atom lifetime studies \cite{maga00, deoliv02, nasc06,marc09,tret09a,feng09,gabb06}) has no ability to parse the zero-Kelvin lifetime from BBR-stimulated depopulation rates, the theoretical calculations do distinguish these effects. Comparison of experiment with theory then becomes dependent on theoretical calculations for the BBR depopulation rates, and as we describe, the works described in \cite{maga00, deoliv02, nasc06, marc09,feng09} all use an inaccurate equation for these rates.

\section{Review of previous theoretical results for Rb Rydberg state lifetimes}\label{rev}

The total radiative decay rate of a Rydberg state at temperature $T$, $\Gamma_{\textrm{\footnotesize{eff}}}(T)$, is related to the radiative lifetime of that state, $\tau_{\textrm{\footnotesize{eff}}}$, by the relation $\Gamma_{\textrm{\footnotesize{eff}}}(T) = 1/\tau_{\textrm{\footnotesize{eff}}}$. It is convenient to consider two radiative depopulation effects that constitute the total radiative decay rate separately. The first effect is the decay rate in the absence of black body radiation, $\Gamma_0 = 1/\tau_0$, where $\tau_0$ will be referred to throughout this paper as the ``zero-Kelvin lifetime''. The second effect is the stimulated black body depopulation rate due to radiation with a characteristic temperature $T$, $\Gamma_{\textrm{\footnotesize{BBR}}}(T)$. The total radiative decay rate is the sum of these two effects:

\begin{equation}
\Gamma_{\textrm{\footnotesize{eff}}}(T) = \Gamma_0 + \Gamma_{\textrm{\footnotesize{BBR}}}(T). \label{gammadef}
\end{equation}

\noindent
The zero-Kelvin vacuum radiation field that produces $\Gamma_0$ generally results in transitions from the target state to the lowest energy state accessible by an electric dipole transition. On the other hand, black body radiation stimulates electric dipole transitions from the target to both higher and lower energy states, and the transitions with the highest probability are those with the smallest change in $n$.

\subsection{Calculations of zero-Kelvin lifetimes}\label{zero}

There have been a number of Rydberg state lifetime calculations for alkali metals, but until very recently, there was no theoretical analysis that was directly relevant to the present work on Rb. There are three principal reasons for this.  First, the theoretical studies were confined to low-$n$ values, for instance Theodosiou's work \cite{theo84} was limited to $n \le 20$ for Rb, and the study of Gounand \cite{goun79} considered only $n \le 28$, though that of He {\it et al.} \cite{he90} extended up to $n = 50$. (All three of these theory papers calculated lifetimes for all Group I elements except H and Fr.) Second, while Theodosiou calculated lifetimes for 0 K, 350 K, and 460 K black body radiation fields, both Gounand and He \etal calculated only $\tau_0$, i.e., they ignored stimulated depopulation by BBR. Finally, none of these three theoretical works discussed how their results could be extended to arbitrary-temperature BBR fields. Very recently, however, a group in Novosibirsk \cite{bet09} has calculated both the zero-Kelvin lifetimes and black body depopulation rates (discussed below) in all the alkalis at arbitrary temperature for $10 \lesssim n \le 80$ (their paper gives an extensive discussion of the calculation techniques used in previous studies which we omit here).

It is usual to describe theoretical zero-Kelvin lifetimes using a simple analytical expression of the form

\begin{equation}
\tau_0 = \tau_{\textrm{\footnotesize{s}}} n^\epsilon_{\textrm{\footnotesize{eff}}}, \label{eq25}
\end{equation}

\noindent
where $\tau_{\textrm{\footnotesize{s}}}$ and $\epsilon$ are constants found by fitting the calculated $\tau_0$ values as a function of the effective principal quantum number, $n_{\textrm{\footnotesize{eff}}} = n - \delta_{\ell}$, where $\delta_{\ell}$ is the quantum defect of the state in question \cite{gall94}. Values for $\tau_{\textrm{\footnotesize{s}}}$ and $\epsilon$ reported in \cite{goun79}, \cite{he90}, and \cite{bet09} are given in \tref{tab1}. However, fitting different ranges of $n_{\textrm{\footnotesize{eff}}}$ yields slightly different $\tau_{\textrm{\footnotesize{s}}}$ and $\epsilon$ values. This led He \etal to specify the maximum error for the use of the constants outside the range of validity given in \tref{tab1} ($10 \le n \le 30$), and also to propose an alternative fitting function for the behavior of $\tau_0$ versus $n_{\textrm{\footnotesize{eff}}}$

\begin{equation}
\tau_0 = a_0 + a_1 n_{\textrm{\footnotesize{eff}}} + a_2 n^2_{\textrm{\footnotesize{eff}}} + a_3 n^3_{\textrm{\footnotesize{eff}}}, \label{eq26}
\end{equation}

\noindent
where $a_0$, $a_1$, $a_2$, and $a_3$ are obtained from the fit and are given in table 9 of \cite{he90}. The maximum error when using \eref{eq26} outside the specified range of $10 \le n \le 30$ is significantly lower than for \eref{eq25}.

\begin{table}
\caption{\label{tab1} Values of the parameters $\tau_{\textrm{\footnotesize{s}}}$ and $\epsilon$ in \eref{eq25} obtained from the results of three major theoretical lifetime calculations of $\tau_0$ for Rb Rydberg states. Note that \cite{goun79} and \cite{he90} do not consider the different $j$-values of the $np$ and $nd$ states, while \cite{bet09} calculated the lifetimes and fit parameters for $ns_{1/2}$, $np_{1/2}$, $np_{3/2}$, $nd_{3/2}$ and $nd_{5/2}$ states. For the $np_{3/2}$ states, two sets of values for $\tau_{\textrm{\footnotesize{s}}}$ and $\epsilon$ are given for the work of Beterov \etal \cite{bet09}. The values identified by an asterisk ($\ast$) are as published in \cite{bet09}, while those identified by a dagger ($\dag$) are corrected values, and are to appear as an erratum \cite{bet09b} (see section \ref{least}).} 

\begin{indented}
\lineup
\item[]\begin{tabular}{ll*{2}{l}l}
\br 
Reference & State & $\tau_{\textrm{\footnotesize{s}}}$ & $\epsilon$ & Validity range \cr                             

& & (ns) & & \cr

\mr
Gounand \cite{goun79} & $ns_{1/2}$ & 1.43 & 2.94 & $10 \le n \le 28$ \cr 
& $np_{}$ & 2.76 & 3.02 \cr
& $nd_{}$ & 2.09 & 2.85 & \cr

He \etal \cite{he90} & $ns_{1/2}$ & 1.25 & 2.99 & $10 \le n \le 30$ \cr
& $np_{}$  & 2.65 & 2.95  \cr
& $nd_{}$ & 1.65 & 2.84 &  \cr

Beterov \etal \cite{bet09} & $ns_{1/2}$ & 1.368 & 3.0008 & $10 \lesssim n \le 80$ \cr
& $np_{1/2}$  & 2.4360 & 2.9989 \cr
& $np_{3/2}$  & 2.5341$^\ast$ & 3.0019$^\ast$ \cr
& $np_{3/2}$  & 2.2214$^\dag$ & 3.0026$^\dag$ \cr
& $nd_{3/2}$  & 1.0761 & 2.9898 \cr
& $nd_{5/2}$ & 1.0687 & 2.9897  \cr

\br
\end{tabular}
\end{indented}
\end{table}

\subsection{Calculations of black body depopulation rates}\label{bbr}

The BBR depopulation rate, ${\Gamma_{\textrm{\footnotesize{BBR}}}(T)}$ has been calculated at different levels of precision. Cooke and Gallagher \cite{cook80} and others \cite{maga00, deoliv02, nasc06, marc09, feng09} have used the following approximation:

\begin{equation}
{{\Gamma^{App}_{\textrm{\footnotesize{BBR}}}}} = {\frac {4 \alpha^3 k_{\textrm{\footnotesize{B}}} T} {3 {n^2_{\textrm{\footnotesize{eff}}}}}}, \label{bbapprox}
\end{equation}

\noindent
where atomic units (A.U.) are assumed, and the symbols have their standard meanings. While the accuracy of the approximation improves as $n_{\textrm{\footnotesize{eff}}}$ increases, the work of Farley and Wing \cite{farl81} reveals that it overestimates the black body decay rates by $\approx$20\% for $ns$ states, $\approx$30\% for $np$ states, and $\approx$40\% for $nd$ states in Rb at $n = 30$. Unfortunately, Farley and Wing's work on BBR decay rates of hydrogen, helium, and the alkalis, extends only up to $n=30$.

Recently, Beterov \etal \cite{bet09} calculated $\Gamma_{\textrm{\footnotesize{BBR}}}(T)$ numerically for alkali atom Rydberg states in the range $10 \lesssim n \le 80$ at several temperatures, and found a semi-empirical analytical function that accurately fits their results in the form

\begin{equation}
\Gamma^{Bet}_{\textrm{\footnotesize{BBR}}}(T) = \frac {A}{n^D_{\textrm{\footnotesize{eff}}}} \ \frac {2.14 \times 10^{10}}{\exp [315780 \times B / (n^C_{\textrm{\footnotesize{eff}}} T)] - 1} \ \textrm{[s}^{-1}\textrm{]}, \label{betbbr}
\end{equation}

\noindent
where the constants $A$, $B$, $C$, and $D$ are found from the fit and are given in table I in \cite{bet09}.

\section{Experimental}\label{exp}

\subsection{General principle of the experiment}\label{general}
\begin{figure*}
\centerline{\resizebox{0.5\textwidth}{!}{\includegraphics{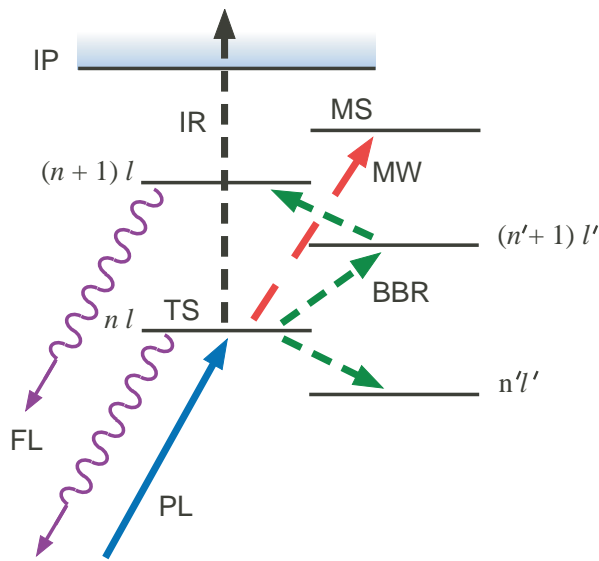}}}
\caption{Diagram illustrating the various techniques that have been used to measure Rydberg state lifetimes. The experiment begins when a target state (TS) is excited by a short laser pulse (PL) at $t=0$. The target state population may then be monitored by SFI with a voltage pulse at a known time delay after $t=0$, and detecting the resulting ions or electrons. Alternatively fluorescence (FL) from radiative decay of the target state can be observed. Another method that has been used is to photo-ionize the target state with a short infrared laser pulse (IR), again detecting the ions or electrons so created. Transitions stimulated by black body photons (BBR) populate states close in energy to the target during the delay between excitation and detection (see text for the consequences of BBR on lifetime measurements). Our experiment uses a frequency-tuned mm-wave pulse (MW) to drive a fixed fraction of the target state population to a higher ``monitor state'' (MS), which is then field ionized. (The atom ionization potential is represented by the line labelled IP.)}

\label{scheme}       

\end{figure*}

The goal of the experiment is to measure the lifetimes of specific target Rydberg states. To do this, one must monitor the population of the target state as a function of time delay after excitation. Additionally, the target state excitation needs to happen on a time scale that is short relative its radiative lifetime. (Figure \ref{scheme} illustrates the excitation process, and the techniques that have been used to monitor the decay of the target state population, which are described below.)

In most Rydberg state lifetime measurements, excitation is done using a pulsed laser ($\lesssim$10 ns pulse length) to excite the target state, and this is the method which we use. Three methods have been used to measure the target state population time-dependence. Specifically, they are: (a) detection of fluorescence from one of the target state radiative decays \cite{goun76,lund76,hugo78,goun80,mare80}; (b) selective field ionization using a voltage pulse \cite{maga00,deoliv02,nasc06,tret09a,marc09,feng09}; and (c) photoionization using a mid-infrared laser pulse \cite{gabb06}. 

All three of the detection methods described above are in some way challenged when high-$n$ Rydberg states are being measured, due to population redistribution caused by BBR. Depopulation of the target state by BBR contributes to the state lifetime at any temperature above zero Kelvin; however, BBR selectively populates states that are close in energy to the target \cite{cook80}, and the method by which the target state population is being monitored needs to reject the signal from these nearby states. For fluorescence detection, a spectrometer may pass only photons originating from the target state. The necessary spectrometer resolution increases for high $n$-states since the energy spacing between adjacent states becomes smaller and thus the photon energies from states with $(n-1)\ell$, $n\ell$, $(n+1)\ell$ (where $n\ell$ is the target) become hard to distinguish. In principle, SFI with a $\sim$ 1 $\mu$s rise-time voltage pulse may be used to distinguish nearby states, provided that electrons (and not ions) are detected. In atomic units, the field required to ionize a state with effective principal quantum number $n_{\textrm{\footnotesize{eff}}}$ is $1/(16 n^4_{\textrm{\footnotesize{eff}}})$ \cite{gall94}. The time-of-flight of the electrons to the detector is much less than 1 $\mu$s, and so they arrive at the detector at essentially the same time that the electric field of the voltage pulse becomes large enough to adiabatically ionize the Rydberg state. Electrons from field ionization of states with different $n_{\textrm{\footnotesize{eff}}}$ arrive at the detector at different times, and the original state can, in principle, be determined. However, states with similar $n_{\textrm{\footnotesize{eff}}}$ but different $\ell$ have different field ionization threshold behaviours \cite{jeys80}, and unambiguous correlation of electrons arriving at the detector at a certain time with a specific $n\ell$ state is not possible at high $n$ \cite{tate07}. Finally, an intense mid-infrared laser pulse with frequency $\nu$ (e.g., from the 10.6 $\mu$m CO$_2$ laser used in \cite{gabb06}) will photoionize almost any state closer to the ionization limit than $h \nu$. This technique therefore has no state specificity.

The detection strategy of our experiment is shown schematically in \fref{scheme}, and is a modification of the SFI detection technique. We avoid the limitation in the SFI method by applying a short ($\sim 1 \ \mu$s) pulse of mm-waves just before the SFI pulse to ``tag'' the target state. We use this ability to identify the part of the SFI electron signal that unambiguously mirrors the target state population, and rejects signal from nearby states populated by BBR. Briefly, a short (0.5 - 5 $\mu$s), frequency tuned, mm-wave pulse is used to excite a fixed fraction of the target state population to a higher ``monitor'' state at a well-determined delay after the target state is populated. We then field ionize the monitor state with a voltage pulse at a fixed time after the end of the mm-wave pulse. The SFI pulse amplitude is set so that the monitor state is ionized, but not the target, and we gate and detect the electron signal at a specific time-of-flight. As described below, these two techniques (mm-wave excitation and gating the electron signal) provide the necessary discrimination against signals originating from states close in energy to the target. 

We now describe the four main parts of the apparatus. Specifically, these are the magneto-optical trap, the pulsed laser system, the mm-wave apparatus, and the timing and detection electronics.

\subsection{Magneto-optical trap}\label{mot}

Our apparatus is basically a vapour-cell MOT \cite{mon90} with internal field plates for applying electric fields and a micro-channel plate detector (MCP) \cite{and98,mou98}. A vapour of rubidium atoms is contained in a cylindrical stainless steel vacuum chamber of diameter 150 mm and height 200 mm. The chamber is pumped by a 20 l/s ion pump to a base pressure of less than $1 \times 10^{-9}$ torr. There is ample optical access for the trapping laser beams, as well as for pulsed lasers and mm-wave radiation. There is a magnetic field gradient of up to 0.2 T/m at the centre of the chamber, created by a pair of air-cooled anti-Helmholtz coils outside the chamber. The source of rubidium atoms is a side-arm heated to 60 - 80$^\circ$ C.

The Rb atoms are cooled by three orthogonal, retro-reflected laser beams of the appropriate circular polarizations. These beams are derived from a single
external cavity diode laser \cite{arn98} that is locked to the appropriate hyperfine component of the $5s_{1/2}$ $\rightarrow$ $5p_{3/2}$ transition in $^{85}$Rb. This light is amplified (after suitable isolation) by an injection-locked non-cavity diode laser, expanded to a diameter of 10 mm, and split into the three orthogonal beams that are introduced into the vacuum chamber. Each trapping laser beam has a power of $\approx 20$ mW. A second amplified external-cavity laser system is used as a repumper. The trapped atoms are observed using a video camera, and we measure the diameter of the cloud of atoms by imaging the fluorescence from the atoms onto a linear diode array. We determine the number of trapped atoms by measuring the fluorescence emitted into a known solid angle using a sensitive optical power meter. Using this apparatus, we trap up to $1.2 \times 10^8$ $^{85}$Rb atoms in a spherical volume of diameter $\approx$1.5 mm (FWHM inferred from diode array signal). The maximum cold-atom number density is therefore $\sim 5 \times 10^{10}$ cm$^{-3}$, though we typically run our experiments at number densities which are at least ten times lower than this to avoid collisional depopulation of the target state. Rubidium atoms trapped by this method typically have temperatures of $\sim 100 \ \mu$K, corresponding to atom velocities of $\lesssim$ 0.2 m/s. 

The atoms are trapped midway between two parallel field plates made from high-transparency stainless steel mesh, separated by 18.3 mm. Excitation of $5p_{3/2}$ atoms to a Rydberg state is achieved using a tunable pulsed laser (see section~\ref{laser}), and this state is field ionized at a specified delay after excitation using a $\sim 1000$ V, 1 $\mu$s rise time voltage pulse applied to the parallel meshes. The resulting electrons are detected using the MCP. 

\subsection{Pulsed laser}\label{laser}

We use a blue ($\approx$480 nm) pulsed laser to excite $ns$ and $nd$ Rydberg states of $^{85}$Rb starting from the $5p \ ^2P_{3/2}$ state. The $np$ states are also accessible provided a small (1 - 5 V/cm) 1 $\mu$s-long electric field pulse is applied simultaneously with the laser pulse. (The $5p_{3/2}$ state is the upper state of the MOT trapping and repump transitions, and close to 50\% of the cold Rb atoms in the MOT are in this state.) For $5p_{3/2}$ $\rightarrow$ $n\ell$ transitions, the Doppler widths of optical transitions for 100 $\mu$K Rb atoms are less than 1 MHz. The spectral width of such a transition is thus limited by the $5p_{3/2}$ state's 5 MHz natural width, since the $n\ell$ Rydberg states have much longer lifetimes than the $5p_{3/2}$ state. For this reason, many groups excite cold Rydberg states using continuous-wave (cw) lasers \cite{lieb05,sing04,vogt06,bend09} because the 1 MHz linewidth available from commercial cw dye, diode, and titanium-sapphire lasers is ideal for exciting and performing spectroscopy on cold Rydberg atoms. 

On the other hand, it is difficult to make the short pulse of light ($\lesssim100$ ns long) needed to excite the target state for a radiative lifetime measurement from a cw laser source. Instead, we used a variation of a Nd:YAG pumped pulsed dye laser. However, when exciting cold Rydberg atoms using a $\sim$10 ns pulsed dye laser (for instance a H\"ansch- \cite{hans72} or a Littmann-type \cite{litt78} dye laser) shot-to-shot fluctuations in the target state population are very large ($\pm$100\% is not unusual). This is due mainly to the rapidly changing spectral overlap between the longitudinal modes of the dye laser and the spectral width of the atomic transition from one laser shot to the next, an issue discussed fully in \cite{li04}. To reduce the shot-to-shot excitation efficiency variation, we seed a Nd:YAG pumped dye amplifier with light from an external cavity diode laser (ECDL). The output of this amplifier is frequency-doubled to produce narrow-bandwidth pulses of 480 nm light \cite{li05, li06}. 

\begin{figure*}
\centerline{\resizebox{0.8\textwidth}{!}{\includegraphics{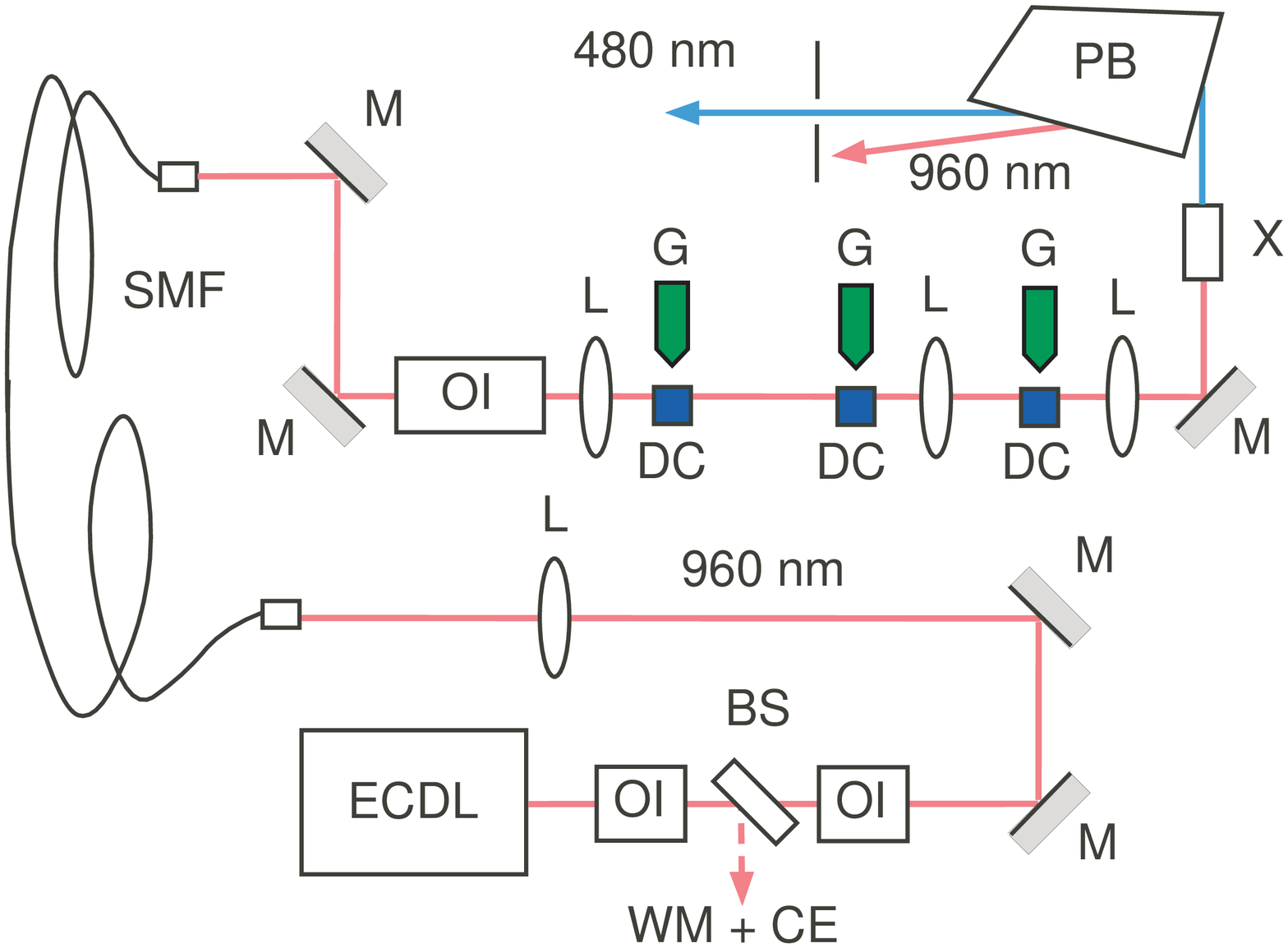}}}
\caption{Schematic diagram of the laser system: ECDL, 960 nm external cavity diode laser; OI, optical isolator; SMF, single-mode fibre; L, lens; WM, wavemeter; CE, confocal etalon; DC, dye cuvette with LDS 925 dye; M, mirror; PB, Pellin-Broca prism; G, 532 nm pump light from Nd:YAG laser; X, KNbO$_3$ frequency doubling crystal; BS, beam splitter.}

\label{laserdiag}       

\end{figure*}

A schematic diagram of our laser system appears in \fref{laserdiag}. The ECDL (Sacher Lasertechnik TEC-520-0960-100) wavelength was tunable between 967 nm and 958 nm, which, when frequency doubled, can excite $5p_{3/2}$ $\rightarrow$ $n\ell$ where 25 $\le$ $n$ $\le$ $\infty$. Light from the laser is passed through two Faraday isolators ($\ge$72 dB of isolation), and a small part is split off and directed towards a wavemeter (Burleigh WA-1500) and a 0.5 meter confocal etalon. For optimal stability, the ECDL is located on an optical table, but the dye amplifier is much less sensitive to acoustic and mechanical noise caused by the Nd:YAG pump laser. We therefore convey the 960 nm light from the ECDL through a fibre to the table where the dye amplifier is located. We use three flowing dye amplifier cells containing LDS925 dye. The first cell has a concentration of 0.06 g/l of LDS 925 dissolved in 15\% propylene carbonate and 85\% ethylene glycol. The latter two cells have a dye concentration of 0.12 g/l. The cells are pumped by 532 nm light from a 20 Hz Nd:YAG laser (Continuum Surelite-II), and the energy/pulse directed to each cell (from the first stage to the third stage) are 20 mJ, 10 mJ, and 20 mJ respectively. We then frequency double the amplified 960 nm light in a $3 \times 3 \times 5$ mm$^3$ KNbO$_3$ crystal, and separate out the 480 nm light using a Pellin-Broca prism. Using this apparatus and seeding the amplifier with 10 - 15 mW of 960 nm light (after the fibre and isolators), we obtain as much as 200 $\mu$J/pulse with new dye, but more typically 50 - 100 $\mu$J/pulse. The spectral width of the 480 nm light is Fourier transform limited by the 10 ns Nd:YAG pulse to approximately 200 MHz. (We can easily resolve the Rb $32d_{3/2}$ and $32d_{5/2}$ states which are separated by 370 MHz.) The $\approx$480 nm light from the pulsed laser is directed into the centre of the MOT chamber via a 1-metre focal length lens which is defocussed from the centre of the trap so that all the cold atoms are illuminated.

Using this system, the shot-to-shot variation in the Rydberg signal is $<\pm$10\%. (This variation could perhaps be improved by pumping the amplifier with an injection-seeded Nd:YAG laser, but the present stability is adequate for our purposes.) To excite a particular $5p_{3/2}$ $\rightarrow$ $n\ell$ transition, the ECDL needs to be tuned to within 100 MHz of the correct transition frequency. The necessary ECDL frequencies were calculated using known physical data for Rb \cite{arim77, lee78, ye96, thib81, li03, han06}, and the laser frequency was measured to a precision of $\pm$30 MHz using the wavemeter.

\subsection{The millimetre-wave system}\label{millimetre}

\begin{figure*}
\centerline{\resizebox{0.8\textwidth}{!}{\includegraphics{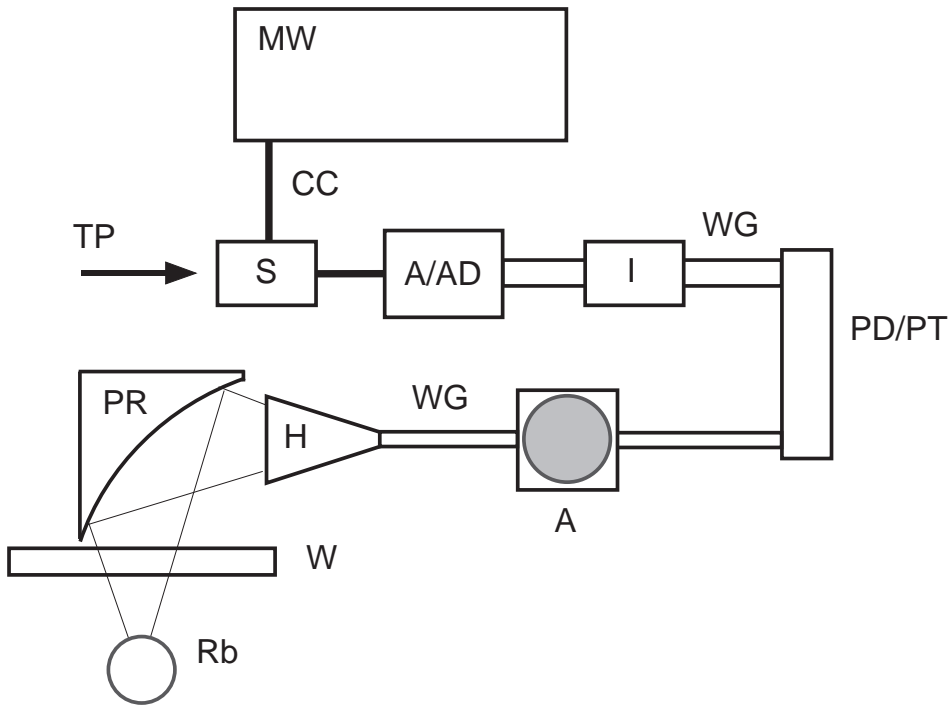}}}
\caption{Schematic diagram of mm-wave system: MW, microwave sweep oscillator; CC, coaxial cable (SMA); S, microwave switch; TP, trigger pulse; I, isolator; A/AD, amplifier or active doubler; WG, waveguide; PD/PT, passive doubler or passive tripler; A, attenuator; H, horn; PR, paraboloid reflector; W, MOT vacuum chamber window; Rb, cold Rb atoms.}

\label{micro}       

\end{figure*}

A schematic diagram of our mm-wave source can be seen in \fref{micro} \cite{li03, li05, li06, han06}. It can generate some 10 mW of radiation in the range 20 - 110 GHz and, at lower power levels, frequencies of up to 200 GHz. The source is based on a 10 MHz - 26.5 GHz sweep oscillator capable of producing 200 mW from 10 - 20 GHz, and 10 mW from 20 - 26.5 GHz (Hewlett-Packard HP83630). The sweep oscillator has a resolution of 1 Hz. The output from the sweep oscillator is switched to make pulses of arbitrary length (General Microwave F9114A; 0.5 - 5 $\mu$s pulse lengths are typical in this experiment), and then amplified and/or frequency multiplied using active or passive GaAs diode doublers or triplers (Pacific Millimeter and Narda) \cite{li03}. We use four different combinations of multipliers and amplifiers to cover the 20 - 110 GHz range, each optimized to a particular frequency interval. For the range 20 - 40 GHz, only an active doubler (Narda DBS-2640X220) is used; for 50 - 75 GHz and 75 - 110 GHz, the active doubler is followed by a passive doubler (Pacific Millimeter V2W0) and a passive tripler (Pacific Millimeter W3W0) respectively. For 40 - 54 GHz, we amplify the sweep oscillator output (Narda DBS-0618N223), and then use a passive tripler (Pacific Millimeter U3). Once the appropriate frequency is generated, the mm-waves are directed into the MOT chamber using a horn and gold-coated off-axis paraboloid reflector through a 150 mm diameter viewport. 

Various mm-wave transitions were used to populate the monitor state from the target. For the $ns_{1/2}$ lifetime measurements, we used either a $ns_{1/2}$ $\rightarrow$ $np_{3/2}$ single-photon transition or a $ns_{1/2}$ $\rightarrow$ $(n+1)s_{1/2}$ two-photon transition. For the $np_{3/2}$ measurements, either $np_{3/2}$ $\rightarrow$ $(n+1)s_{1/2}$ (one-photon) or $np_{3/2}$ $\rightarrow$ $(n+1)p_{3/2}$ (two-photon) transitions were used. Finally, for the $nd_{5/2}$ measurements, either $nd_{5/2}$ $\rightarrow$ $(n-2)f_{7/2}$ (one-photon) or $nd_{5/2}$ $\rightarrow$ $(n+1)d_{5/2}$ (two-photon) transitions were employed.

\subsection{Timing and detection electronics, data acquisition and analysis}\label{synch}

\begin{figure*}
\centerline{\resizebox{0.8\textwidth}{!}{\includegraphics{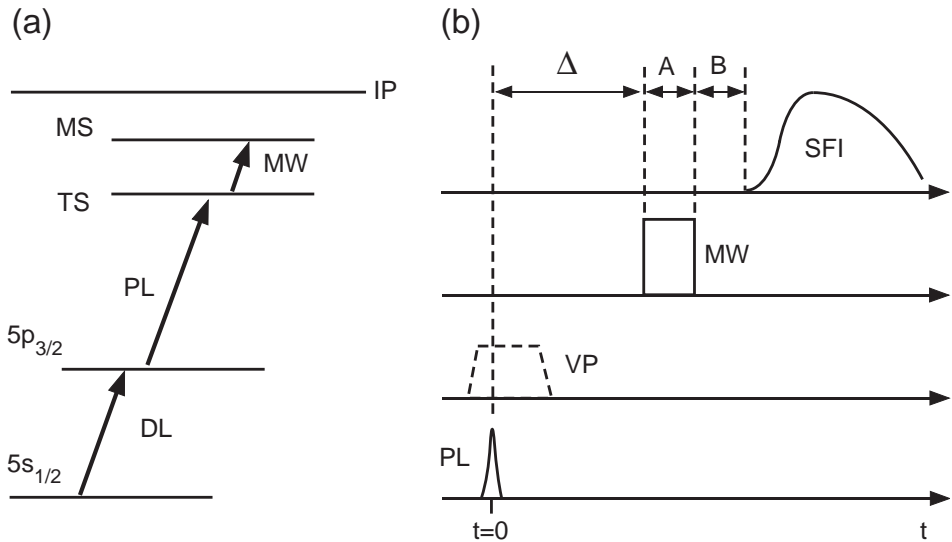}}}
\caption{Schematic diagram of timing of experiment. In (a), a partial energy diagram of $^{85}$Rb is shown, along with the excitation scheme of the target state (whose lifetime is to be determined): DL, MOT diode lasers (trapping and repump lasers) drive the $5s_{1/2}$ $\rightarrow$ $5p_{3/2}$ transition; PL, pulsed laser (480 nm, 10 ns); MW, mm-wave pulse; TS, target state, MS, monitor state; IP, ionization potential. In (b), the timing sequence is shown: t, time; VP, optional voltage pulse (maximum electric field $\approx$5 V/cm, length $\approx$1 $\mu$s) allowing excitation of $np$ Rydberg states from $5p_{3/2}$; SFI, selective field ionization pulse ionizes monitor state but not target state; $\Delta$, delay of mm-wave pulse after 480 nm laser pulse; A, mm-wave pulse length; B delay of SFI pulse after mm-waves; all other symbols as in (a).}

\label{time}       

\end{figure*}

The 480 nm laser pulse, the mm-wave pulse, SFI pulse, and the detection electronics (boxcar amplifier and digital oscilloscope) are all synchronized using two digital delay generators (DDG; Stanford Research Systems DDG-535). One DDG is used a master oscillator, and is internally triggered at 20 Hz. The master DDG triggers the flashlamp and Q-switch of the Nd:YAG laser used to pump the dye amplifier for the ECDL (see section \ref{laser}). The master DDG also triggers an analog pulse generator which is used to apply a small ($\le 5$ V/cm), 1 $\mu$s long pulse to the Rb atoms just before the arrival of the 480 nm laser pulse when $np$ states are being excited. The rest of the experiment is synchronized by the second DDG, which is triggered (at time $t=0$) by the master DDG. The second DDG provides three output pulses that synchronize the mm-waves and the detection apparatus, as diagrammed in \fref{time}. One pulse controls the microwave switch in the mm-wave system (see section \ref{millimetre}). This pulse is delayed by $\Delta$ after $t=0$, and has a duration $A$ of between 0.5 $\mu$s and 5 $\mu$s, depending on the state being measured (we ensure that $A \ll \tau_{\textrm{\footnotesize{eff}}}$). At a fixed delay of $B = 2 \ \mu$s after the microwave pulse is switched off, the second pulse from the DDG triggers the SFI pulse. The SFI pulse is applied to one of the two parallel meshes surrounding the cold atoms in the MOT (see section \ref{mot}), and ionizes the monitor state (and all higher energy states), but does not ionize the target state. The resulting electrons are pushed towards the MCP. The output of the MCP is amplified and passed to a digital oscilloscope and a boxcar amplifier. (The third pulse from the DDG triggers the digital oscilloscope and boxcar amplifier 1 $\mu$s before the SFI pulse is triggered.) The boxcar amplifier is set to acquire the part of the electron time-of-flight signal corresponding to field ionization of the monitor state.

The second DDG, the microwave sweep oscillator, and the boxcar amplifier are controlled by a computer via a GPIB interface using a program written in the \textit{LabVIEW} environment (National Instruments). The program controls the microwave oscillator output frequency and power level, and can turn the microwave output on and off. The program also sets the mm-wave pulse length $A$, and steps the delay $\Delta$ from an initial value to a maximum value in 20 equal steps. Typically, the initial value of $\Delta$ is between 1 and 5 $\mu$s, and the maximum value of $\Delta$ is between 60 and 200 $\mu$s later than the initial value, depending on the state being measured. At each delay step $\Delta$, \textit{LabVIEW} acquires the time-gated MCP signal from the monitor state for a predetermined number of laser pulses (usually 10) when the microwave oscillator is on (resulting in a measurement $S(\Delta)_{\textrm{\footnotesize{on}}}$) and when it is off ($S(\Delta)_{\textrm{\footnotesize{off}}}$). The means and standard deviations of the means of $S(\Delta)_{\textrm{\footnotesize{on}}}$ and $S(\Delta)_{\textrm{\footnotesize{off}}}$ for the 10 laser shots are calculated. The delay is then stepped back to ``zero'' (this means the initial value of $\Delta$), and the corresponding means and standard deviations of the means are calculated for $S(\Delta=0)_{\textrm{\footnotesize{on}}}$ and $S(\Delta=0)_{\textrm{\footnotesize{off}}}$ over 10 laser shots each. For this $\Delta$, \textit{LabVIEW} calculates the fraction $F$

\begin{equation}
F = \frac {S(\Delta)_{\textrm{\footnotesize{on}}} - S(\Delta)_{\textrm{\footnotesize{off}}}} {S(\Delta=0)_{\textrm{\footnotesize{on}}} - S(\Delta=0)_{\textrm{\footnotesize{off}}}}, \label{f}
\end{equation}

\noindent
along with its standard deviation ($\sigma_F$). Subtracting the signal acquired when the mm-waves are off allows us to reject spurious signals from electrons originating from SFI of states other than the monitor state, and from BBR stimulated population of the monitor state from the target state, neither of which is a constant fraction of the target state population. In addition, repetitively stepping $\Delta$ back to zero allows us to normalize against slow variations in the target state population, due, for instance, to changes in the MOT atom density and variations in the 480 nm laser pulse energy and frequency. \textit{LabVIEW} then steps $\Delta$ to the next delay value, and the process is repeated until we have 21 $F$ and $\sigma_F$ values for 21 different $\Delta$ values. The $F$ versus $\Delta$ curve is then plotted and fitted to find the decay lifetime using \textit{Igor Pro} (Wavemetrics). The uncertainty in the $\Delta$ measurement is negligible due to the high precision of the DDG's used in the timing electronics ($\sigma_{\Delta} \lesssim 10$ ns), while the uncertainty in each $F$ value is $\sigma_F$, and the $\sigma_F$ values are used to weight the decay curve fit.

We measure the lifetime of each state between 5 and 40 times. From each state's set of lifetime measurements, we calculate the mean of the set, which is our experimental lifetime determination, and also the standard deviation of the mean. This standard deviation of the mean is combined in quadrature with the mean uncertainty arising from the fitting process to find the uncertainty in the lifetime measurement. The resulting experimental uncertainty is between 2\% and 8\% of the measured lifetime, with around 5\% being typical.

\section{Results and discussion}\label{res}

In the following sections, we present our results, and compare them with other recent experimental results. We also extract the zero-Kelvin lifetimes and decay rates from our results using Beterov's equation for $\Gamma_{\textrm{\footnotesize{BBR}}}(T)$ (see \eref{betbbr} above), and compare them with previous theoretical results. Finally, we discuss the impact of the BBR spectral temperature on the inferred zero-Kelvin lifetimes.

\subsection{Experimental results}

\begin{figure*}
\centerline{\resizebox{0.6\textwidth}{!}{\includegraphics{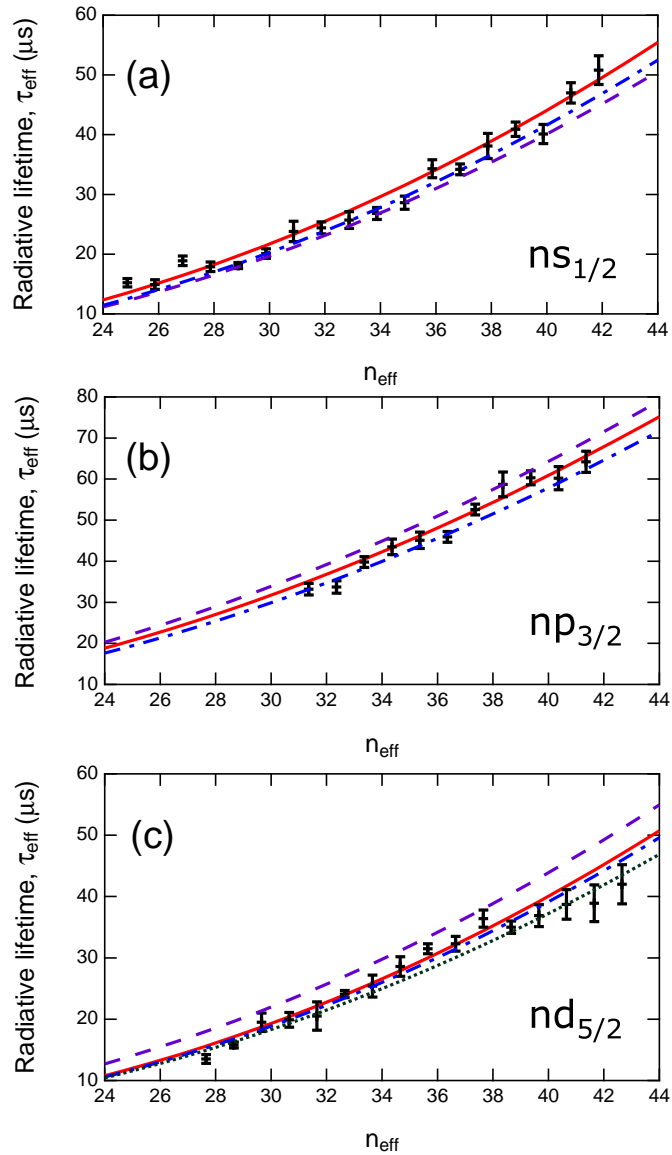}}}
\caption{Our results for the radiative lifetimes, $\tau_{\textrm{\footnotesize{eff}}}^{Exp}(T)$ of (a), the Rb $ns_{1/2}$ ($28 \le n \le 45$) states; (b), the $np_{3/2}$ ($34 \le n \le 44$) states; and (c), the $nd_{5/2}$ ($29 \le n \le 44$) states, at a nominal radiation temperature of $T=300$ K, plotted versus the effective principal quantum number, $n_{\textrm{\footnotesize{eff}}}$. The graphs also show theoretical predictions based on \eref{gammadef}, \eref{betbbr}, \eref{eq25}, and \eref{eq26}, using parameter values reported in a number of different papers. All the models use the BBR rates of Beterov \etal (equation (14) in \cite{bet09}); the solid red line (\full) uses values for $\tau_{\textrm{\footnotesize{s}}}$ and $\epsilon$ from Beterov \etal  in \eref{eq25}; the purple dashed (\longbroken) \ line uses Gounand's values for $\tau_{\textrm{\footnotesize{s}}}$ and $\epsilon$ \cite{goun79}; the blue dash-dot-dash (\chain) \ line uses the values of He \etal for $a_0....a_3$ in \eref{eq26} \cite{he90}. The predictions using the values of He \etal for $\tau_{\textrm{\footnotesize{s}}}$ and $\epsilon$ in \eref{eq25} are essentially indistinguishable from those found using their values for $a_0....a_3$ in \eref{eq26}, except for the $nd_{5/2}$ curve, which is shown as a green dotted (\dotted) line.}

\label{results1}       

\end{figure*}

Our experimental lifetimes for the Rb $ns_{1/2}$, $np_{3/2}$ and $nd_{5/2}$ states are listed in \tref{tab5}, and plotted in \fref{results1}. We obtained lifetimes for all the $ns_{1/2}$ states in the range $28 \le n \le 45$ (18 states), for $np_{3/2}$ states in the range $34 \le n \le 44$ (11 states), and for $nd_{5/2}$ states in the range $29 \le n \le 44$ (16 states). This range of states is limited by several factors. At low $n$, the energy spacing of adjacent levels exceeds the maximum mm-wave photon energy available from our apparatus (around 200 GHz), even when two-photon transitions are used. At high $n$, collisions become an issue due to long range dipole-dipole interactions \cite{gall94,and98,mou98,li04,li05,li06}. Though we took extensive measures to work at low background gas pressure and low trap density, we noticed that the lifetime measurements tended towards a limit as $n$ increased above 45-50, an effect that we attributed to collisions. (This is apparent even though our measurements for these states were made with the Rb reservoir valved off from the MOT chamber, i.e., we trapped from the residual Rb atoms in the vacuum chamber.) Another issue that came into play above $n \approx 45$ was the effect of microwaves ``leaking through'' the switch that synchronized the mm-waves with the pulsed laser. As $n$ increases, Rydberg states become more and more sensitive to microwave radiation, and the small amount of leak-through radiation was enough to affect the measured lifetime. For $nd$ states with $n \ge 39$, the bandwidth of the pulsed laser which excites the target state at $t=0$ ($\approx 200$ MHz) exceeds the separations of the $j=3/2$ and $j=5/2$ sub-levels, which is close to 200 MHz at $n=39$ and approximately 120 MHz at $n=44$ \cite{li03}. These $nd$ state lifetime measurements may therefore be subject to some averaging over the different $j$-levels. The mm-wave source that was used to populate the monitor state has a resolution that is more than enough to distinguish the $nd_{3/2}$ $\rightarrow$ $(n+1)d_{3/2}$ two-photon transitions from the $nd_{5/2}$ $\rightarrow$ $(n+1)d_{5/2}$ two-photon transitions that were used for these states, which are separated by between 20 MHz and 10 MHz in the range $39 \le n \le 44$. However, it is possible that power broadening may result in some $j=3/2$ character being present in the nominally ``pure'' $j=5/2$ monitor state signal for these target states. On the other hand, the calculations of Beterov \etal indicate that the lifetimes of the two $45d_j$ sub-levels differ by less than 0.2 $\mu$s at 300 K (see table VII in \cite{bet09}), so the impact of any inadvertent averaging on our results is probably negligible.

Two other features of our experiment have a potential to impact our lifetime measurements. First, our experiments were performed without switching the anti-Helmholtz coils off for a short time in synchronism with the laser pulse. Hence, the MOT magnetic field was on at all times, which could lead to state mixing due to the Zeeman interaction. Given the spatial size of the atoms in the MOT ($\approx 1$ mm), and the fact that the atoms were trapped at the zero-point of a magnetic field with a gradient of $\le 0.2$ T/m, the maximum field magnitude experience by the Rydberg atoms was $\approx 10^{-4}$ T. Such small fields are well below the Paschen-Back regime, which is reached around $10^{-3}$ - $10^{-2}$ T for the $np_{1/2}$ - $np_{3/2}$ and $nd_{3/2}$ - $nd_{5/2}$ fine structure splittings at $n = 50$ \cite{gall94}. Hence, both $\ell$ and $j$ are good quantum numbers, and additionally, any splitting of the different $m_j$ levels of a given $j$ state are much smaller than the pulsed laser linewidth. In other words, we can be certain which $n\ell_j$ state we are exciting, and there is no $\ell$-mixing of states with different lifetimes. 

The second issue is that there is a small residual electric field in the interaction region, which could lead to Stark mixing of different $\ell$ states. The electric field is due primarily to the bias voltage on the MCP, which (at the time of these experiments) was not housed in a grounded enclosure, though it was separated from the interaction region by the meshes described in section \ref{mot}. We have used this same apparatus to study ultra-cold plasmas (similar to those described in \cite{li04} and \cite{kill99}). In the plasma experiments we found it necessary to apply a nulling field of no more than 1 V/cm to maximize the plasma lifetime. However, this field was not corrected in the lifetime measurements described in this paper, and we therefore estimate that the Rydberg atoms experienced a static electric field of $\le 1$ V/cm. The Stark interaction mixes states with orbital angular momentum $\ell$ with states having $\ell^\prime = \ell \pm 1$ (which themselves mix with states $\ell^\prime \pm 1$). However, $m_\ell$ (or, in atoms with large fine structure splittings like Rb, $m_j$) remains a good quantum number. This mixing obviously has the potential to impact our lifetime measurements, given the generally much longer lifetimes of the $np$ states than of the $ns$ and $nd$ states (whose lifetimes are similar) at comparable $n_{\textrm{\footnotesize{eff}}}$.  

To estimate the extent to which our measurements are an average over Stark-mixed $n\ell$ states, we have calculated the amount of mixing that could result from an electric field using a program that diagonalizes the Stark energy matrix \cite{zimm79}. For a given electric field, the mixing is stronger as $n$ increases, so we calculated the mixing at the highest $n_{\textrm{\footnotesize{eff}}}$ for any state whose lifetime was measured in our experiments ($44d_{5/2}$, corresponding to $n_{\textrm{\footnotesize{eff}}} \approx 43$). Additionally, we assumed an electric field of 2 V/cm, i.e., twice the field that was present in our experiments. We found that in this regime, the admixture of states with a different orbital angular momentum quantum number $\ell$ was negligible. For instance, the $44d_{5/2}$ $|m_j| = 1/2$ state is 1.57\% (probability) $\ell = 1$, 97.10\% $\ell = 2$, and 1.29\% $\ell = 3$, while the $46s_{1/2}$ $|m_j| = 1/2$ state is 98.95\% $\ell = 0$, and 1.05\% $\ell = 1$. We looked at all the possible $|m_j|$ eigenstates in the $46s_{1/2}$, $45p_{1/2}$, $45p_{3/2}$, $44d_{3/2}$ and $44d_{5/2}$ states in a 2 V/cm field, and found that none had less than 97.1\% (by probability) of its zero-field $\ell$ eigenstate content, and the maximum amount of a state with a different $\ell$ content was the 1.57\% $p$-character in the $44d_{5/2}$ $|m_j| = 1/2$ state. Such small admixtures do not impact the measured state lifetimes at a level that can be distinguished in this experiment. 

We therefore conclude that the residual electric and magnetic fields have no impact on our lifetime measurements. (The $\le 5$ V/cm voltage pulse which is applied simultaneously with the laser when $np$ states are to be excited decays to zero over $\approx 1 \ \mu$s. Hence, the pulse switches off slowly enough that the Stark state excited adiabatically couples to $\ell = 1$ in zero field and introduces no $\ell$-mixing \cite{free76}.)

\begin{table}
\caption{\label{tab5} Our experimentally determined radiative lifetimes for the $ns_{1/2}$, $np_{3/2}$ and $nd_{5/2}$ Rydberg states of Rb at a temperature of 300 K. For the $nd$ state lifetimes denoted by the asterisk, the separation of the $nd$ $j=3/2$ and $j=5/2$ states is less than the bandwidth of the pulsed laser used to populate the target state at $t=0$. The measured lifetime is therefore likely subject to some degree of averaging over the two $j$ states.} 

\begin{indented}
\lineup
\item[]\begin{tabular}{*{4}{l}}
\br                              
State&$ns_{1/2}$&$np_{3/2}$&$nd_{5/2}$ \cr
{$n$}& {($\mu$s)} & {($\mu$s)} & {($\mu$s)} \cr

\mr
28&$15.2 \pm 0.7$& & \cr
29&$14.9 \pm 0.8$& &$13.5 \pm 0.7$\cr 
30&$18.9 \pm 0.8$& &$15.8 \pm 0.5$\cr 
31&$17.9 \pm 0.8$& &$19.5 \pm 1.5$\cr 
32&$18.1 \pm 0.5$& &$19.9 \pm 1.2$\cr
33&$20.1 \pm 0.8$& &$20.5 \pm 2.3$\cr 
34&$23.8 \pm 1.7$&$33.2 \pm 1.4$&$24.1 \pm 0.6$\cr 
35&$24.4 \pm 1.0$&$33.7 \pm 1.5$&$25.4 \pm 1.8$\cr 
36&$25.7 \pm 1.4$&$39.8 \pm 1.3$&$28.6 \pm 1.6$\cr 
37&$26.8 \pm 1.0$&$43.5 \pm 1.9$&$31.5 \pm 0.8$\cr 
38&$28.6 \pm 1.1$&$45.1 \pm 2.0$&$32.3 \pm 1.2$ \cr 
39&$34.3 \pm 1.5$&$45.9 \pm 1.3$&$36.4 \pm 1.4^\ast$\cr 
40&$34.2 \pm 0.9$&$52.6 \pm 1.3$&$35.0 \pm 1.0^\ast$\cr 
41&$38.1 \pm 2.1$&$58.7 \pm 3.0$&$36.9 \pm 1.8^\ast$\cr 
42&$40.9 \pm 1.2$&$60.3 \pm 1.7$&$38.7 \pm 2.4^\ast$\cr 
43&$40.1 \pm 1.6$&$60.2 \pm 2.8$&$38.9 \pm 3.0^\ast$\cr  
44&$47.0 \pm 1.7$&$64.2 \pm 2.6$&$42.0 \pm 3.2^\ast$\cr 
45&$50.8 \pm 2.4$& & \cr 
\br
\end{tabular}
\end{indented}
\end{table}

Our measurements roughly correspond to those made using a slightly different method by a group at Universidade de S\~{a}o Paulo, Brazil \cite{deoliv02, nasc06}. Their full set of results for $ns$ ($28 \le n \le 44$) and $nd$ ($27 \le n \le 42$) states is given in \cite{nasc06} and \cite{marc09}, and their results for the $np$ ($33 \le n \le 42$) in \cite{deoliv02} (their laser excitation method prevented isolation of the different $j$-states for $np$ and $nd$). The $np$ lifetimes given in \cite{deoliv02} were erroneously reported to be twice the actual values, as they pointed out in the reference section of their subsequent paper \cite{nasc06}, and their $np$ lifetime results were afterwards reported correctly in \cite{marc09}. (This group's experimental results for the $ns$, $np$, and $nd$ lifetimes are also reproduced in full in figure 6 of the paper by Beterov \etal \cite{bet09}.) In general, our $s$- and $d$-state results are higher than the S\~{a}o Paulo group's measurements by an average of 7-8\%. While some values differ by 20-30\% for low $n$, there are negative as well as positive differences between the data sets. Furthermore, our $p$-state results are all lower than the previous measurements by between 5\% and 16\%. However, the difference between the two data sets is not large, and almost all the lifetimes agree within three times the combined experimental uncertainty. We conclude that while there seem to be systematic differences between our results and those of the S\~{a}o Paulo group, these differences are comparable with the measurement uncertainties. Tretyakov \etal have also measured the lifetime of the $37p$ state of Rb using a MOT source, obtaining $\tau_{\textrm{\footnotesize}} = 38 \pm 3$ $\mu$s, which is lower than our value of $43.5 \pm 1.9$ $\mu$s \cite{tret09a}.

In a recent paper reporting the first experimental observation of a new class of Rydberg molecules, lifetime measurements were used to distinguish signals from atomic and molecular species. Specifically, Bendkowsky \etal observed $^{87}$Rb$_2$ dimers comprising one atom in the $5s_{1/2}$ ground state, and the other atom in a $ns_{1/2}$ Rydberg state, where $34 \le n \le 40$ \cite{bend09}. The molecular resonances were observed as satellites to atomic resonances when a tunable 480 nm cw laser was scanned over the $5p_{3/2}$ $\rightarrow$ $ns_{1/2}$ transition, and the resonances were observed using SFI followed by detection of the resulting ions with an MCP. The atomic sample had a temperature of 3.5 $\mu$K and a density of $1.5 \times 10^{13}$ cm$^{-3}$, i.e., similar conditions to those used to create a quantum-degenerate gas of Rb. Bendkowsky \etal report a number of experimental observations that strongly justify their claim that they observe new molecular species, including measurements of lifetimes for the different spectral features. While their method was able to distinguish different decay times for the atomic and molecular states, their measured lifetimes of the atomic states are very different from our own (and those of the other experimental works discussed in the previous paragraph), ranging from 65 $\pm$ 9 $\mu$s for $35s$ to 57 $\pm$ 5 $\mu$s for $37s$. These differences are surprising, even when it is considered that in \cite{bend09}, the detection of positive ions meant that the observed signal was insensitive to redistribution by BBR so that many Rydberg states likely contributed to it. On the other hand, the experiments reported in \cite{bend09} were carried out with MOT atom densities $\sim 10^4$ times larger than in our own study and those reported in \cite{maga00,deoliv02,nasc06,marc09,tret09a,gabb06,feng09}. Indeed, it is unclear that an atomic property like a radiative lifetime can be measured under these conditions, given the occurrence of many-body effects at much lower densities \cite{and98,mou98,tong04,sing04,lieb05,vogt06}. It is, however, curious that in this instance, any such effects (or other mechanism) cause longer state lifetimes to be observed for dense samples than for individual atoms.

\subsection{Comparison with theory}

We considered three different means of comparison between our data and existing theoretical results. First, we estimated which theoretical result was most compatible with our data set. Second, we combined our data with data from previous low-$n$ measurements to find experimental values for $\tau_{\textrm{\footnotesize{s}}}$ and $\epsilon$ in \eref{eq25}. Finally, we analyzed the sensitivity of these parameters to our imperfect knowledge of the BBR temperature.

\subsubsection{Least-squares comparison with theory}\label{least}

We synthesized theoretical lifetime values based on \eref{gammadef}, \eref{betbbr}, \eref{eq25}, and \eref{eq26}, using parameter values reported in \cite{goun79,he90,bet09} (we used the values denoted by the asterisk ($\ast$) in \tref{tab1} for the $np_{3/2}$ lifetime predictions of Beterov \etal \cite{bet09}). All the models use \eref{betbbr} for the BBR rates from \cite{bet09}. These synthesized lifetime values are shown as smooth curves in \fref{results1}. The behaviour of the zero-Kelvin lifetimes in the work of Gounand and of Beterov \etal \cite{goun79,bet09} is of the form of \eref{eq25}, while He \etal also use this form, as well as that of \eref{eq26} \cite{he90}. (The predictions using values for $\tau_{\textrm{\footnotesize{s}}}$ and $\epsilon$ from He \etal in \eref{eq25} are essentially indistinguishable from those found using their values for $a_0....a_3$ in \eref{eq26}, except for the $nd_{5/2}$ states.)

We used a variant of the ``chi-squared'' statistical parameter to compare our results with the theoretical curves. Specifically, we calculated the values for

\begin{equation}
\chi^2 \equiv \sum_{i=1}^N \ (\tau_{0,i}^{Exp} - \tau_{0,i}^{Th})^2 \ / \ (\sigma_{0,i}^{Exp})^2, \label{chisq1}
\end{equation}

\noindent
where $\tau_{0,i}^{Exp}$ are our measured values, $\tau_{0,i}^{Th}$ the theoretical values, and $\sigma_{0,i}^{Exp}$ our experimental uncertainties. These parameters were calculated for each of the $ns_{1/2}$, $np_{3/2}$, and $nd_{5/2}$ series. For each angular momentum series, $N$ is the number of different state lifetimes that were measured.

For the $ns_{1/2}$ and $np_{3/2}$ states, $\chi^2$ is consistently lower for the results of Beterov \etal \cite{bet09}, although the predictions of He \etal are almost as good, the $\tau_{0}^{Th}$ values predicted by \eref{eq26} being slightly better than those using \eref{eq25} \cite{he90}. (We have been advised by one of the authors of \cite{bet09} that the $\tau_{\textrm{\footnotesize{s}}}$ and $\epsilon$ for the $np_{3/2}$ states are in error \cite{bet09b}. The corrected values are given in \tref{tab1} of the present paper, and are to appear in an erratum to \cite{bet09}. Using the revised values, the lifetime predictions for the $np_{3/2}$ states almost exactly match those of He \etal using either of \eref{eq25} or \eref{eq26}.) For the $nd_{5/2}$ states, the results of He \etal using \eref{eq26} agree with the data slightly better than those of Beterov \etal, with the prediction of He \etal from \eref{eq25} being worse than either. For all of the measurements, Gounand's predictions \cite{goun79} have a much poorer agreement than either of He \etal results or that of Beterov \etal. 

We conclude that our data set is unable to definitively distinguish which of the works of Beterov \etal or He \etal gives better parameters $\tau_{\textrm{\footnotesize{s}}}$ and $\epsilon$ in \eref{eq25}, nor whether \eref{eq25} or \eref{eq26} given by He \etal is more accurate. However, it seems that the accuracy of the theoretical calculations of both of these theoretical papers is better than that of Gounand \cite{goun79}.

\subsubsection{Experimental dependence of $\tau_0$ on $n_{\textrm{\footnotesize{eff}}}$}

We initially tried fitting our data to \eref{eq25}, assuming a radiation temperature of 300 K and the equation for the BBR depopulation rate, \eref{betbbr}, from Beterov \etal. That is, we equated our experimental lifetimes, $\tau_{\textrm{\footnotesize{eff}}}^{Exp}(T)$, with $1/\Gamma_{\textrm{\footnotesize{eff}}} \ (300 \ \textrm{K})$ in \eref{gammadef} assuming that \eref{betbbr} was correct, to find $\Gamma_0$ and hence experimental values for $\tau_0$ (designated $\tau_0^{Exp}$):

\begin{equation}
\Gamma_0^{Exp} = \frac {1}{\tau_0^{Exp}} = \biggl (\frac {1}{\tau_{\textrm{\footnotesize{eff}}}^{Exp}} - \Gamma^{Bet}_{\textrm{\footnotesize{BBR}}}(T) \biggr ). \label{tau0exp}
\end{equation}

\noindent
We then used \eref{eq25} to extract experimental values for $\tau_{\textrm{\footnotesize{s}}}$ and $\epsilon$ (assuming that $T=300$ K), using a least-squares fit. However, since our measurements were taken at relatively high $n$, there is a large uncertainty in the values obtained, $\approx \pm$100\% for $\tau_{\textrm{\footnotesize{s}}}$ and $\approx \pm$10\% for $\epsilon$. We concluded that using only our own lifetime results was insufficient to make a meaningful comparison with theory.

\begin{figure*}
\centerline{\resizebox{0.6\textwidth}{!}{\includegraphics{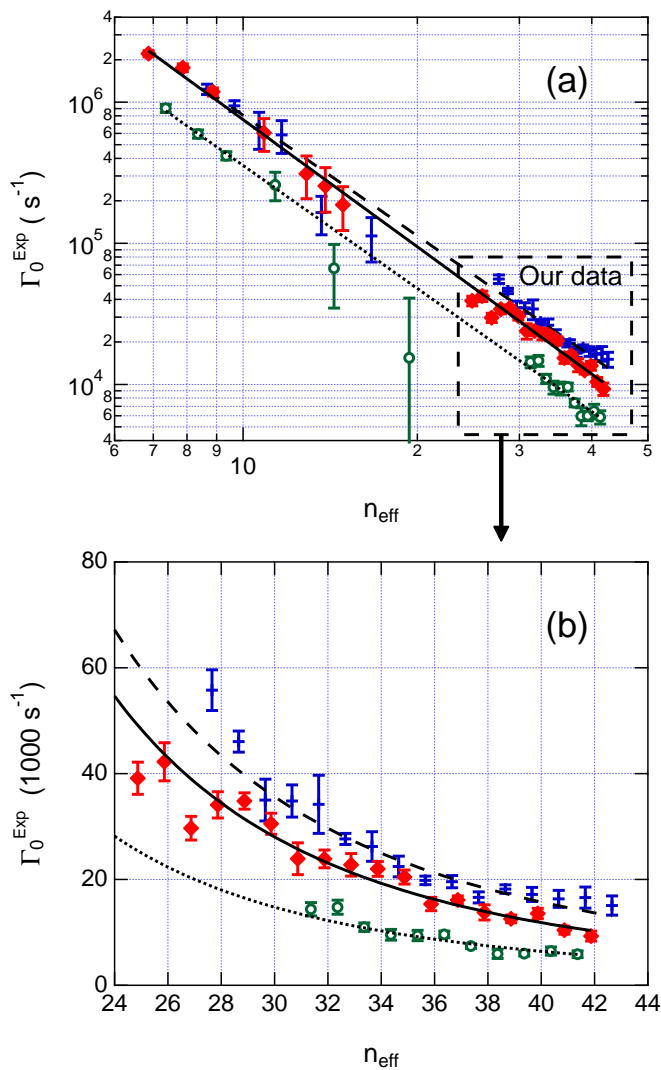}}}
\caption{(a) Log-log plot of an extended set of rubidium Rydberg state zero-Kelvin decay rate measurements, $\Gamma_0^{Exp}$, comprising our results and those of \cite{goun80, mare80, goun76} for the $ns_{1/2}$ (full red diamonds, \fulldiamond), $np_{3/2}$ (open green circles, \opencircle), and $nd_{5/2}$ (blue crosses, $+$) states with $n \ge 10$. The range of $n_{\textrm{\footnotesize{eff}}}$ values is $6 \le n_{\textrm{\footnotesize{eff}}} \le 50$, and the range of $\Gamma_0^{Exp}$ values is $4 \times 10^3 \ \textrm{s}^{-1} \le \Gamma_0^{Exp} \le 4 \times 10^6 \ \textrm{s}^{-1}$. The previous results are the most precise values given in table V of \cite{theo84} that satisfy the criterion that $n \ge 10$. The zero-Kelvin decay rates are obtained from the data using \eref{tau0exp}, assuming that the calculations of Beterov and co-workers for the BBR depopulation rates are correct. The straight lines are fits of the data to \eref{eq25}, i.e., we fit the data to the equation  $\Gamma_0^{Exp} = (1/\tau_{\textrm{\footnotesize{s}}}) \ n_{\textrm{\footnotesize{eff}}}^{-\epsilon}$. The full line (\full) is the fit to the $ns_{1/2}$ data, the dotted line (\dotted) to the $np_{3/2}$ data, and the dashed line (\dashed) to the $nd_{5/2}$ data. The dotted box shows our data, and is expanded in graph (b) below. (b) Inset of data presented in (a) between $24 \le n_{\textrm{\footnotesize{eff}}} \le 44$ showing only our results as a linear-linear plot. The $y$-axis is calibrated in units of 1000 s$^{-1}$. This plot also shows the fit lines to the entire data set shown in (a).}

\label{results2}       

\end{figure*}

To obtain more accurate values for $\tau_{\textrm{\footnotesize{s}}}$ and $\epsilon$, we added data points from previous measurements of Rydberg state lifetimes in Rb. Specifically, we included values for the following states: $ns_{1/2}$ with $n=$ 10, 11, 12, 14, 16, 17, 18; $np_{3/2}$ with $n=$ 10, 11, 12, 14, 17, 22; and $nd_{5/2}$ with $n=$ 10, 11, 12, 13, 15, 18. These experimental values are the most precise measurements given in table V of Theodosiou's paper \cite{theo84}, and were originally reported in \cite{goun76,mare80,goun80}. Using these data points, as well as our own, we plotted the $\Gamma_0^{Exp}$ values obtained using \eref{tau0exp} versus $n_{\textrm{\footnotesize{eff}}}$, and these data are presented in \fref{results2}. 

We excluded previous data for $n < 10$ because the accuracy with which \eref{betbbr} reproduces the numerical result for $\Gamma^{Bet}_{\textrm{\footnotesize{BBR}}}$ is unspecified below $n = 10$ \cite{bet09}. When obtaining the zero-Kelvin decay rates shown in \fref{results2}, we assumed a BBR spectral temperature of exactly 300 K for our own experiment. In fact, this may not be the case, an issue which is considered below in section \ref{temp}. With regard to the previous results shown in \fref{results2}, we have used the estimated BBR temperatures given in those papers. Specifically, in \cite{mare80}, the Rb was contained in a 320 K cell; in \cite{goun76}, the sample was at 460 K; and in \cite{goun80} the Rb was maintained at 520 K. We have assumed these temperatures to be exact, i.e., the uncertainty in the $\Gamma_0^{Exp}$ values is entirely due to the experimental uncertainty of the lifetime measurements, with no contribution from $\Gamma_{\textrm{\footnotesize{BBR}}}(T)$.

We fitted the data presented in \fref{results2} to \eref{eq25} to obtain experimental values for $\tau_{\textrm{\footnotesize{s}}}$ and $\epsilon$ for the $ns_{1/2}$, $np_{3/2}$, and $nd_{5/2}$ states of Rb (see \tref{tab4}). Comparing these experimental results to the theoretical values given in \tref{tab1}, it appears that the dependence of $\tau_0^{Exp}$ on $n_{\textrm{\footnotesize{eff}}}$ is best reproduced by He \etal \cite{he90}, rather than by that of Beterov \etal (in which the parameter $\epsilon$ shows essentially no dependence on $\ell$). The three theoretical papers \cite{goun79, he90, bet09} substantially agree on the $\tau_{\textrm{\footnotesize{s}}}$ values for the $ns_{1/2}$ and $np_j$ series, and our value of $\tau_{\textrm{\footnotesize{s}}}$ for $ns_{1/2}$ is consistent with this finding, although our value for $np_j$ is higher than the theoretical values by $\approx$30\%. For the $nd_j$ data, the $\tau_{\textrm{\footnotesize{s}}}$ value best agrees (within 10\%) with the result from He \etal, being 15\% lower than Gounand's and 70\% higher than that of Beterov \etal.

It is unclear whether any differences of the experimental $\tau_{\textrm{\footnotesize{s}}}$ values from those given in any of the three theoretical papers is meaningful, given the 10\% uncertainty of the experimental values. On the other hand, a preliminary consideration of the experimental $\epsilon$ values obtained from the extended data set seems to favor the predictions of He \etal, rather than those of Gounand or Beterov \etal. However, we believe that any such comparison must be treated with caution, given the uncertainty in our knowledge of the BBR temperature, which we discuss below. 

\begin{table}
\caption{\label{tab4} Values of the parameters $\tau_{\textrm{\footnotesize{s}}}$ and $\epsilon$ in \eref{eq25} obtained by fitting this equation to the data given in \fref{results2}.} 

\begin{indented}
\lineup
\item[]\begin{tabular}{*{3}{l}}
\br                              
State&$\tau_{\textrm{\footnotesize{s}}}$ &$\epsilon$ \cr
& (ns) & \cr
\mr
$ns_{1/2}$&$1.4 \pm 0.1$&$2.99 \pm 0.03$\cr
$np_{3/2}$&$3.5 \pm 0.4$&$2.90 \pm 0.03$\cr 
$nd_{5/2}$&$1.8 \pm 0.3$&$2.84 \pm 0.04$\cr 
\br
\end{tabular}
\end{indented}
\end{table}

Nascimento \etal \cite{nasc06} found $\tau_{\textrm{\footnotesize{s}}} = 1.43 \pm 0.05$ ns and $\epsilon = 2.94 \pm 0.03$ for the Rb $ns_{1/2}$ states in the range $28 \le n \le 44$, and $\tau_{\textrm{\footnotesize{s}}} = 1.90 \pm 0.07$ ns and $\epsilon = 2.83 \pm 0.04$ for the Rb $nd_{j}$ states in the range $27 \le n \le 42$ (these values are also reported in \cite{marc09}). These results agree quite well with our findings shown in \tref{tab4}, and with theory \cite{goun79,he90,bet09}. However, as one of us has observed \cite{tate07}, any agreement of the results in \cite{nasc06} with ours, or with the theoretical values, is apparently accidental, since Nascimento \etal used the approximate equation \eref{bbapprox} for BBR depopulation rates to find the zero-Kelvin lifetimes from their experimental data. As noted above, the approximate equation overestimates the BBR depopulation rates by between 20\% and 40\% at $n=30$ in Rb. (A very recent experimental paper on cesium Rydberg state lifetimes also uses the approximation for the BBR rates when comparing results with theory \cite{feng09}, as does \cite{marc09} when comparing the $\tau_{\textrm{\footnotesize{s}}}$ and $\epsilon$ values for Rb $np$ states with theory.)

\subsubsection{Effect of BBR temperature}\label{temp}

The ability of BBR to significantly affect state lifetimes was investigated in several studies in the late 1970's \cite{gall79,duca79,beit79}. We have not attempted to measure or control the BBR temperature with any high degree of precision in our experiment, and this approach is similar to other recent Rydberg lifetime experiments \cite{maga00,deoliv02,nasc06,marc09,tret09a,gabb06,feng09}. Our lab is stabilized at 295 K; however, numerous pieces of equipment become warm during use, and, as pointed out by Spencer \etal \cite{spen82a}, it is very difficult to shield any experiment from BBR from external sources. Most of the BBR absorbed by atoms in our apparatus will come from the walls of the vacuum chamber, and will have a radiation temperature close to 300 K. However, the atoms will also be illuminated by BBR from the Rb side-arm, and this source was heated to between 330 and 350 K. In other words, there is no single temperature that characterizes the BBR field, since the vacuum chamber is not in thermal equilibrium. 

To investigate the potential impact on the values of $\Gamma_0$ and $\tau_0$ that we inferred from our experimental data, we performed a simple analysis using expressions from Beterov \etal for $\tau_0$ (\eref{eq25} and the values for $\tau_{\textrm{\footnotesize{s}}}$ and $\epsilon$ given at the bottom of \tref{tab1}) and $\Gamma_{\textrm{\footnotesize{BBR}}}^{Bet}$ (\eref{betbbr} and the values for $A$, $B$, $C$, and $D$ given in table I in \cite{bet09}). We make the gross assumption that \eref{eq25} and \eref{betbbr}, when used with parameter values  from Beterov \etal, exactly predict the result of an experiment carried out on atoms illuminated with BBR with characteristic temperature $T$. In this situation, the lifetime values measured, $\tau_{\textrm{\footnotesize{eff}}}^{Exp}$, would be

\begin{equation}
\frac {1}{\tau_{\textrm{\footnotesize{eff}}}^{Exp}(T)} = \frac {1}{\tau_0} + \Gamma^{Bet}_{\textrm{\footnotesize{BBR}}}(T). \label{modeleq1}
\end{equation}

\noindent
On the other hand, if these values of $\tau_{\textrm{\footnotesize{eff}}}^{Exp}$ are used to infer values for the zero-Kelvin lifetime, $\tau_0^{Inf}$, assuming that the BBR temperature is 300 K, we have

\begin{equation}
\frac {1}{\tau_{\textrm{\footnotesize{eff}}}^{Exp}(T)} = \frac {1}{\tau_0^{Inf}} + \Gamma^{Bet}_{\textrm{\footnotesize{BBR}}}(300 \ \textrm{K}). \label{modeleq2}
\end{equation}

\noindent
Obviously, if $T = 300$ K, fitting the values of $\tau_0^{Inf}$ to \eref{eq25} would yield the original values for $\tau_{\textrm{\footnotesize{s}}}$ and $\epsilon$. The question we want to investigate is, what happens to the values of these parameters obtained if $T \ne 300$ K? We consider two situations: $T = 270$ K, and $T = 330$K, and made another gross assumption, namely that the ``experimental'' values of $\tau_0^{Inf}$ have uncertainties that are a fixed percentage of their absolute values. (If the data points are unweighted, the low-$n$ lifetimes have a much smaller impact on the fit than the high-$n$ lifetimes.) Finally, we assume that we ``measured'' lifetimes of only the states shown in \fref{results2}. 

We did this analysis for the $ns$, $np$, and $nd$ series, but we will discuss only the $np$ results. Because the $np$ states have longer zero-Kelvin lifetimes, BBR has a bigger impact on the effective $np$ lifetimes, and therefore a wrong temperature assignment has a bigger impact on the inferred zero-Kelvin lifetime. We found that while a difference of 10\% between the actual value of $T$ and its assumed value (300 K) changed $\epsilon$ by only 3\%, from approximately 3.0 to 2.9 or 3.1, the value of $\tau_{\textrm{\footnotesize{s}}}$ changed by about 15\%, from 2.5 ns to 2.1 ns or 2.9 ns. (The range of these values does not correspond to the $\tau_{\textrm{\footnotesize{s}}}$ value for the $np_{3/2}$ states given in \tref{tab4} since we are starting with theoretical values from \cite{bet09}, not our own experimental data.) As expected, the effect on the $ns$ and $nd$ results was smaller. 

We conclude that the ability of experimental measurements of Rydberg state lifetimes to inform the results of theoretical calculations is substantially dependent on accurate knowledge of the BBR field temperature. This is an issue that can only be resolved if BBR depopulation rates of the Rydberg states can be measured under the same experimental conditions as those of the lifetime measurements. It is unclear, however, whether the BBR transition rates can be measured with sufficient precision to yield zero-Kelvin lifetimes with small enough uncertainties for useful comparison with theory. The two studies that have specifically addressed this question, those of Spencer \etal \cite{spen82a} and Galvez \etal \cite{galv99}, did not measure the BBR transfer rates explicitly. Rather, they showed that their models (based on theoretical rates) were consistent with the observed time-evolution of population over a range of Rydberg states \cite{galv99}, or that the observed variation of the radiative lifetime of a state with temperature was in agreement with calculations \cite{spen82a}.

\section{Conclusion}\label{conc}

We have measured the radiative lifetimes of high-$n$ Rydberg states of rubidium using cold atoms in a MOT. The slow velocities of these atoms allow them to be observed for long enough that lifetimes of states up to at least $n=50$ may be measured. Our experiment has a significant improvement over earlier work that reduces the impact of a possible source of systematic error. Specifically, we use mm-waves to ``tag'' a particular state so that the detected signal is an accurate mirror of the target state population. This eliminates ambiguities in the signal caused by BBR stimulated population transfer to nearby Rydberg states.

Our results are consistent with other recent experimental studies \cite{deoliv02,nasc06,marc09}, and also agree well with several theoretical papers \cite{goun79,he90,bet09}. However, we find that the ability of our data to distinguish the merits of the three different theoretical studies is limited by a lack of experimental data of BBR depopulation rates in rubidium, and by our imperfect knowledge of the BBR spectral temperature. On the other hand, we have used available theoretical information on BBR rates to obtain the highest possible accuracy for our values for the zero-Kelvin lifetimes of Rb Rydberg states.

\vfill\eject
\ack

We acknowledge many conversations with Mike Noel, and especially Tom Gallagher, Wenhui Li, and Paul Tanner, who gave invaluable assistance with the design of the mm-wave spectrometer and the narrow-bandwidth pulsed laser system. Charlie Conover provided much assistance on calculating Stark-mixing effects due to residual electric fields, as well as in many other matters. We appreciate the willingness of Phillip Gould to share with us plans of the mount for the high-transparency field ionization meshes. Ryan Jennerich and Greg Foltz assisted with the construction of the MOT used in this experiment. This work has been supported by NSF (grant numbers OIA-9873763, PHY-0140430, and PHY-0652842) and by Colby College via the Division of Natural Sciences Grant Program. Tamas Juhasz and Cristian Vesa acknowledge support from Colby College in the form of summer research assistantships, and Roy Wilson is grateful to the NSF for a supplemental summer research assistantship (grant number PHYS-0734190).

\bibliography{JPB_Bib_102109}

\providecommand{\newblock}{}
\begin{thebibliography}{10}
\expandafter\ifx\csname url\endcsname\relax
  \def\url#1{{\tt #1}}\fi
\expandafter\ifx\csname urlprefix\endcsname\relax\def\urlprefix{URL }\fi
\providecommand{\eprint}[2][]{\url{#2}}

\bibitem{hans72}
H\"{a}nsch T~W 1972 {\em Appl. Opt.\/} {\bf 11} 895

\bibitem{litt78}
Littman M~G and Metcalf H~J 1978 {\em Appl.\ Opt.\/} {\bf 17} 2224

\bibitem{goun79}
Gounand F 1979 {\em \JP (Paris)\/} {\bf 40} 457

\bibitem{theo84}
Theodosiou C~E 1984 {\em \PR A\/} {\bf 30} 2881

\bibitem{he90}
He X, Li B, Chen A and Zhang C 1990 {\em \JPB\/} {\bf 23} 661

\bibitem{bet09}
Beterov I~I, Ryabtsev I~I, Tretyakov D~B and Entin V~M 2009 {\em \PR A\/} {\bf
  79} 052504

\bibitem{gall79}
Gallagher T~F and Cooke W~E 1979 {\em \PRL\/} {\bf 42} 835

\bibitem{duca79}
Ducas T~W, Spencer W~P, Vaidyanathan A~G, Hamilton W~H and Kleppner D 1979 {\em
  Appl. Phys. Lett.\/} {\bf 35} 382

\bibitem{beit79}
Beiting E~J, Hildebrandt G~F, Kellert F~G, Foltz G~W, Smith K~A, Dunning F~B
  and Stebbings R~F 1979 {\em \JCP\/} {\bf 70} 3551

\bibitem{spen82a}
Spencer W~P, Vaidyanathan A~G, Kleppner D and Ducas T~W 1982 {\em \PR A\/} {\bf
  25} 380

\bibitem{farl81}
Farley J~W and Wing W~H 1981 {\em \PR A\/} {\bf 23} 2397

\bibitem{galv99}
Galvez E~J, Lewis J~R, Chaudhuri B, Rasweiler J~J, Latvakoski H, De~Zela F,
  Massoni E and Castillo H 1995 {\em \PR A\/} {\bf 51} 4010

\bibitem{and98}
Anderson W~R, Veale J~R and Gallagher T~F 1998 {\em \PRL\/} {\bf 80} 249

\bibitem{mou98}
Mourachko I, Comparat D, de~Tomasi F, Fioretti A, Nosbaum P, Akulin V~M and
  Pillet P 1998 {\em \PRL\/} {\bf 80} 253

\bibitem{tong04}
Tong D, Farooqi S~M, Stanojevic J, Krishnan S, Zhang Y~P, C\^ot\'e R, Eyler E~E
  and Gould P~L 2004 {\em \PRL\/} {\bf 93} 063001

\bibitem{sing04}
Singer K, Reetz-Lamour M, Amthor T, Marcassa L~G and Weidem\"uller M 2004 {\em
  \PRL\/} {\bf 93} 163001

\bibitem{lieb05}
Cubel-Liebisch T, Reinhard A, Berman P~R and Raithel G 2005 {\em \PRL\/} {\bf
  95} 253002

\bibitem{vogt06}
Vogt T, Viteau M, Zhao J, Chotia A, Comparat D and Pillet P 2006 {\em \PRL\/}
  {\bf 97} 083003

\bibitem{stan06}
Stanojevic J, C\^ot\'e R, Tong D, Farooqi S~M, Eyler E~E and Gould P~L 2006
  {\em \EJP D\/} {\bf 40} 3

\bibitem{bend09}
Bendkowsky V, Butscher B, Nipper J, Shaffer J~P, Low R and Pfau T 2009 {\em
  Nature\/} {\bf 458} 1005

\bibitem{li04}
Li W, Noel M~W, Robinson M~P, Tanner P~J, Gallagher T~F, Comparat D,
  Laburthe-Tolra B, Vogt T, Zahzam N, Vanhaecke N, Pillet P and Tate D~A 2004
  {\em \PR A\/} {\bf 70} 042713

\bibitem{li05}
Li W, Tanner P~J and Gallagher T~F 2005 {\em \PRL\/} {\bf 94} 173001

\bibitem{li06}
Li W, Tanner P~J, Jamil Y and Gallagher T~F 2006 {\em \EJP D\/} {\bf 40} 27

\bibitem{li03}
Li W, Mourachko I, Noel M~W and Gallagher T~F 2003 {\em \PR A\/} {\bf 67}
  052502

\bibitem{han06}
Han J, Jamil Y, Norum D~V~L, Tanner P~J and Gallagher T~F 2006 {\em \PR A\/}
  {\bf 74} 054502

\bibitem{maga00}
{Magalh\~aes} K~M~F, de~Oliveira A~L, Zanon R~A~D~S, Bagnato V~S and Marcassa
  L~G 2000 {\em Opt.\ Commun.\/} {\bf 184} 385

\bibitem{deoliv02}
de~Oliveira A~L, Mancini M~W, Bagnato V~S and Marcassa L~G 2002 {\em \PR A\/}
  {\bf 65} 031401(R)

\bibitem{nasc06}
Nascimento V~A, Caliri L~L, de~Oliveira A~L, Bagnato V~S and Marcassa L~G 2006
  {\em \PR A\/} {\bf 74} 054501

\bibitem{marc09}
Marcassa L~G 2009 {\em \PS\/} {\bf T134} 014011

\bibitem{gabb06}
Gabbanini C 2006 {\em Spectrochim.\ Acta,\ Part\ B\/} {\bf 61} 196

\bibitem{tret09a}
Tretyakov D~B, Beterov I~I, Entin V~M, Ryabtsev I~I and Chapovsky P~L 2009 {\em
  JETP\/} {\bf 108} 374

\bibitem{feng09}
Feng Z~G, Zhang L~J, Zhao J~M, Li C~Y and Jia S~T 2009 {\em \JPB\/} {\bf 42}
  145303

\bibitem{tate07}
Tate D~A 2007 {\em \PR A\/} {\bf 75} 066502

\bibitem{cali07}
Caliri L~L and Marcassa L~G 2007 {\em \PR A\/} {\bf 75} 066503

\bibitem{gall94}
Gallagher T~F 1994 {\em Rydberg Atoms\/} (Cambridge, UK: Cambridge University
  Press)

\bibitem{bet09b}
Beterov I~I 2009 Private communication

\bibitem{cook80}
Cooke W~E and Gallagher T~F 1980 {\em \PR A\/} {\bf 21} 588

\bibitem{goun76}
Gounand F, Fournier P, Cuvellier J and Berlande J 1976 {\em Phys. Lett. A\/}
  {\bf 59} 23

\bibitem{lund76}
Lundberg H and Svanberg S 1976 {\em Phys. Lett. A\/} {\bf 56} 31

\bibitem{hugo78}
Hugon M, Gounand F and Fournier P~R 1978 {\em \JPB\/} {\bf 11} L605

\bibitem{goun80}
Gounand F, Hugon M and Fournier P 1980 {\em \JP (Paris)\/} {\bf 41} 119

\bibitem{mare80}
Marek J and Munster P 1980 {\em \JPB\/} {\bf 13} 1731

\bibitem{jeys80}
Jeys T~H, Foltz G~W, Smith K~A, Beiting E~J, Kellert F~G, Dunning F~B and
  Stebbings R~F 1980 {\em \PRL\/} {\bf 44} 390

\bibitem{mon90}
Monroe C, Swann W, Robinson H and Wieman C 1990 {\em \PRL\/} {\bf 65} 1571

\bibitem{arn98}
Arnold A~S, Wilson J~S and Boshier M~G 1998 {\em Rev.\ Sci.\ Instrum.\/} {\bf
  69} 1236

\bibitem{arim77}
Arimondo E, Inguscio M and Violino P 1977 {\em \RMP\/} {\bf 49} 31

\bibitem{lee78}
Lee S~A, Helmcke J, Hall J~L and Stoicheff B~P 1978 {\em Opt. Lett.\/} {\bf 3}
  141

\bibitem{ye96}
Ye J, Swartz S, Jungner P and Hall J~L 1996 {\em Opt. Lett.\/} {\bf 21} 1280

\bibitem{thib81}
Thibault C, Touchard F, B\"uttgenbach S, Klapisch R, de~Saint~Simon M, Duong
  H~T, Jacquinot P, Juncar P, Liberman S, Pillet P, Pinard J, Vialle J~L,
  Pesnelle A and Huber G 1981 {\em \PR C\/} {\bf 23} 2720

\bibitem{kill99}
Killian T~C, Kulin S, Bergeson S~D, Orozco L~A, Orzel C and Rolston S~L 1999
  {\em \PRL\/} {\bf 83} 4776

\bibitem{zimm79}
Zimmerman M~L, Littman M~G, Kash M~M and Kleppner D 1979 {\em Phys. Rev. A\/}
  {\bf 20} 2251

\bibitem{free76}
Freeman R~R and Kleppner D 1976 {\em \PR A\/} {\bf 14} 1614

\end{thebibliography}

\end{document}